\documentclass{iaus}
\usepackage{graphicx}

\title[Helium star donor channel to SNe Ia and their surviving companions] 
{Helium star donor channel to type Ia supernovae and their surviving companion stars}

\author[Wang \& Han]   
{Bo Wang$^{1,2}$ \and Zhanwen Han$^{1,2}$}

\affiliation{$^1$National Astronomical Observatories/Yunnan Observatory, CAS, Kunming, 650011, China (email: {\tt wangbo@ynao.ac.cn})
\\[\affilskip]
$^2$Key Laboratory for the Structure and Evolution of Celestial Objects, CAS, Kunming, 650011, China}

\pubyear{2011}
\volume{281}  
\pagerange{1--4}
\jname{Binary Paths to the Explosions of type Ia Supernovae}
\editors{R. Di Stefano \& M. Orio, eds.}
\begin{document}

\maketitle

\begin{abstract}
Employing Eggleton's stellar evolution code with the optically thick
wind assumption, we systematically studied the He star donor channel
of type Ia supernovae (SNe Ia), in which a carbon-oxygen white dwarf (WD) accretes material
from a He main-sequence star or a He subgiant to increase its mass
to the Chandrasekhar mass. We mapped out the initial parameters for
producing SNe Ia in the orbital period--secondary mass plane for
various WD masses from this channel. Based on a detailed binary
population synthesis approach, we find that this channel can produce
SNe Ia with short delay times ($\sim$100\,Myr) implied by recent
observations. We obtained many properties of the surviving
companions of this channel after SN explosion, which can be verified
by future observations. We also find that the surviving companions
from the SN explosion scenario have a high spatial velocity
($>$400\,km/s), which could be an alternative origin for
hypervelocity stars (HVSs), especially for HVSs such as US 708.


\end{abstract}

\firstsection 
\section{Introduction}
Type Ia Supernovae (SNe Ia) have been applied
successfully in determining cosmological parameters and are also a key part of our understanding of galactic
chemical evolution. However, there is still no agreement on the nature of
their progenitors. It is generally believed that SNe Ia are
thermonuclear explosions of carbon-oxygen white dwarfs (CO WDs) in
binaries (Nomoto et al. 1997).
Over the past few decades, two families of SN Ia progenitor models
have been proposed, i.e., the double-degenerate (DD) and
single-degenerate (SD) models. Of these two models, the SD model is
widely accepted at present. It is suggested that the DD model, which
involves the merger of two CO WDs (Iben \& Tutukov
1984; Webbink 1984), likely leads to an
accretion-induced collapse rather than to an SN Ia (Nomoto \& Iben 1985). For
the SD model, the companion could be a main-sequence (MS) star or a
slightly evolved subgiant star (WD + MS channel), or a red-giant
star (WD + RG channel) (Hachisu et al.
1996; Li $\&$ van den Heuvel 1997; Han $\&$ Podsiadlowski 2004, 2006; Chen $\&$ Li 2007;
Meng \& Yang 2010; Wang, Li \& Han 2010; Wang \& Han 2010a; Di Stefano et al. 2011).
Meanwhile, a CO WD may also accrete material from a He star to
increase its mass to the Chandrasekhar (Ch) mass, which is known as
the He star donor channel. WD + He star systems are believed to originate from
intermediate mass binaries, which may explain SNe Ia with short
delay times implied by recent observations (Mannucci et al. 2006).
The aim of this contribution is to investigate the He star donor channel and to explore the
properties of the surviving companions after SN explosion.

\section{Binary evolution calculations}
We use Eggleton's stellar evolution code (Eggleton 1971)
to calculate the evolutions of WD + He star systems. The code
has been updated with the latest input physics over the past four
decades. In our calculations, He star models are
composed by He abundance $Y=0.98$ and metallicity $Z=0.02$.
Instead of solving stellar structure equations of a WD, we use the
optically thick wind model (Hachisu et al. 1996) and adopt the prescription
of Kato \& Hachisu (2004) for the mass accumulation efficiency
of He-shell flashes onto the WD. We have calculated about 2600 WD +
He star systems, and obtained a large, dense model grid (see Wang et al.
2009a). In Fig. 1, we show the contours at the onset of
Roche lobe overflow (RLOF) for producing SNe Ia in the $\log P^{\rm i}-M^{\rm i}_2$ plane
for various WD masses, where $P^{\rm i}$ and $M^{\rm i}_2$
are the orbital period and the mass of the He companion star at the
onset of RLOF, respectively. If the
parameters of a WD + He star system at the onset of the RLOF are
located in the contours, an SN Ia is then assumed to be produced.
Thus, these contours can be expediently used in binary population synthesis (BPS) studies to
investigate the He star donor channel.

\begin{figure}[tb]
   \begin{center}
\includegraphics[width=5.4cm,angle=270]{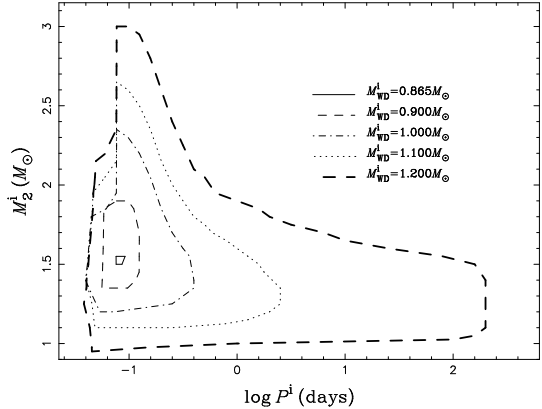}
 \caption{Regions in the orbital
period--secondary mass plane for WD binaries (at the onset of the
RLOF stage) that produce SNe Ia. (From Wang et al.
2009a)}
   \end{center}
\end{figure}

\section{Binary population synthesis}
To obtain SN Ia birthrates and delay times of the He star donor
channel, we performed a series of Monte Carlo simulations in the BPS
study (see
Wang et al. 2009b). In the simulation, by using the Hurley's rapid binary
evolution code (Hurley et al. 2002), we followed the evolution of
$4\times10^{\rm 7}$ sample binaries from the star formation to the
formation of the WD + He star systems based on the SN Ia
production regions (Fig. 1) and three formation channels (see Wang et al.
2009b). Here, we
adopt the standard energy equations to calculate the output of the
common envelop phase.

The simulations give Galactic SN Ia birthrate of $\sim$$0.3\times
10^{-3}\ {\rm yr}^{-1}$ by adopting $Z=0.02$ and ${\rm SFR}=5\,M_{\rm
\odot}{\rm yr}^{-1}$, which is
lower than that inferred observationally (i.e., $3 - 4\times
10^{-3}\ {\rm yr}^{-1}$; Cappellaro and Turatto 1997). This
implies that the He star donor channel is only a subclass of SN Ia
production, and there may be some other channels or mechanisms also
contributing to SNe Ia (see Wang et al. 2010). The simulations also show the evolution of
SN Ia birthrates with time for a single starburst
with a total mass of $10^{11}$$\,M_{\odot}$. We find that
that SN Ia explosions occur between $\sim$45\,Myr and $\sim$
140\,Myr after the starburst, which could explain SNe Ia with short
delay times implied by recent observations (Mannucci et al. 2006).


The companion in the SD model would survive in the SN explosion and
potentially be identifiable (Wang \& Han 2010b). We obtained many
properties of the surviving companions of the He star donor channel
at the moment of SN explosion, e.g., the masses, the orbital velocities, the effective
temperatures, the surface gravities, the orbital period and the
equatorial rotational velocity of companions (see Wang \& Han 2009). These
distributions may be helpful for identifying the surviving
companions of SNe Ia. The simulation also shows the initial
parameters of the primordial binaries and the WD + He star systems
that lead to SNe Ia, which may help to search for potential SN Ia
progenitors. In Fig. 2, we show the current epoch distribution of the spatial
velocity for the surviving companions from this channel (see Wang \& Han 2009).
We see that the surviving companions have high
spatial velocities ($>$400\,km/s), which almost exceed the
gravitational pull of the Galaxy nearby the Sun. Thus, the surviving
companions from the SN explosion scenario could be an alternative
origin for hypervelocity stars (HVSs), which are stars with a
velocity so great that they are able to escape the gravitational
pull of the Galaxy.

\begin{figure}[tb]
   \begin{center}
\includegraphics[width=8.0cm,angle=0]{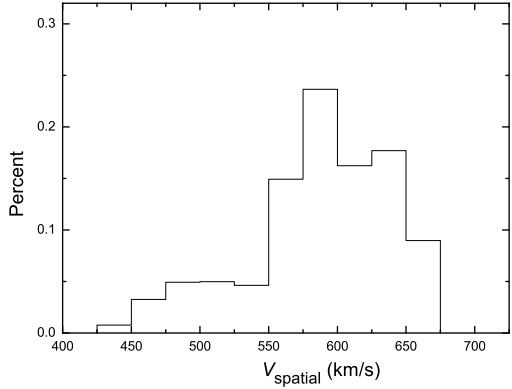}
 \caption{The distribution of the spatial velocity with
 $\alpha_{\rm ce}\lambda=0.5$. (From Wang \& Han 2009)}
    \end{center}
\end{figure}

\section{Discussion}

Based on the optically thick wind assumption, Wang \& Han
(2010c) recently calculated about 10 000 WD + He star systems
and obtained SN Ia production regions of the He star donor channel
with different metallicities. For a constant star-formation galaxy,
they found that SN Ia birthrates increase with metallicity. If a
single starburst is assumed, SNe Ia occur systemically earlier and
the peak value of the birthrate is larger for a high $Z$. We also
note that Liu et al. (2010) recently investigated the effects
of rapid differential rotation on the accreting WD, and found that
the highest mass of the accreting WD at the moment of SN Ia
explosion is 1.81$\,M_{\odot}$ for the He star donor channel, which
may provide a way for the formation of super-Ch mass SNe Ia in the
observations.

Massive WD + He star systems are candidates of SN Ia progenitors. HD
49798/RX J0648.0-4418 is an evidence of the existence of massive WD
+ He star systems. Based on the data from the XMM-Newton satellite,
Mereghetti et al. (2009) recently derived the masses of the
two components. The corresponding masses are
1.50$\pm$0.05$\,M_{\odot}$ for HD 49798 and
1.28$\pm$0.05$\,M_{\odot}$ for the WD. Based on a detailed
binary evolution model, Wang \& Han (2010d) found that the
massive WD can increase its mass to the Ch mass in future evolution. Thus, HD 49798 with its WD companion is a
candidate of SN Ia progenitors. V445 Pup is another candidate of
massive WD + He star system, in which the mass of the WD is more
than 1.35\,$M_{\odot}$, and the mass of the He star is more than
0.85$\,M_{\odot}$ (Kato et al. 2008).


US 708 is an extremely He-rich sdO star in the halo of the Galaxy, with a
heliocentric radial velocity of +$708\pm15$\,${\rm km/s}$ (Hirsch et
al. 2005). We note that the local velocity relative to the
Galatic center may lead to a higher observation velocity for the
surviving companions, but this may also lead to a lower observation
velocity. Considering the local velocity nearby the Sun
($\sim$220\,km/s), we find that $\sim$30\% of the surviving
companions may be observed to have velocity $V>700\,{\rm km/s}$ for
a given SN ejecta velocity 13500\,km/s. In addition, the asymmetric
explosion of SNe Ia may also enhance the velocity of the surviving
companions. Thus, a surviving companion in the He star donor channel
may have a high velocity like US 708. In future investigations, we
will employ the Large sky Area Multi-Object fiber Spectral Telescope
(LAMOST) to search the HVSs originating from the surviving
companions of SNe Ia.

%
This work is supported by the Natural Science Foundation of China under
grant nos. 10821061, 11033008, 2007CB815406 and 2009CB824800.



\begin{thebibliography}{}
\bibitem[Cappellaro \& Turatto (1997)]{ct97}     Cappellaro, E.,  \&  Turatto, M. 1997, in Ruiz-Lapuente P., Cannal R., Isern J., eds, Thermonuclear Supernovae. Kluwer, Dordrecht, P. 77
\bibitem[Chen \& Li (2007)]{che07}               Chen, W.-C., \& Li, X.-D. 2007, ApJ, 658, L51
\bibitem[Di Stefano et al. (2011)]{di11}         Di Stefano,  R., Voss, R.,  \& Claeys, J.S.W. 2011, submitted to ApJ Letters (arXiv:1102.4342)
\bibitem[Eggleton (1971)]{egg71}                 Eggleton, P.P. 1971, MNRAS, 151, 351
\bibitem[Hachisu et al. (1996)]{hac96}           Hachisu, I., Kato, M., \& Nomoto, K. 1996, ApJ, 470, L97
\bibitem[Han \& Podsiadlowski (2004)]{han04}     Han, Z., \& Podsiadlowski, Ph. 2004, MNRAS, 350, 1301
\bibitem[Han \& Podsiadlowski (2006)]{han06}     Han, Z., \& Podsiadlowski, Ph. 2006, MNRAS, 368, 1095
\bibitem[Hirsch et al. (2005)]{hir05}            Hirsch, H.A., Heber, U., O'Toole, S.J., \& Bresolin, F. 2005, A\&A, 444, L61
\bibitem[Hurley et al. (2002)]{hur02}            Hurley, J.R., Tout, C.A., \& Pols, O.R. 2002, MNRAS, 329, 897
\bibitem[Iben \& Tutukov (1984)]{it84}           Iben, I., \& Tutukov, A.V. 1984, ApJS, 54, 335
\bibitem[Kato \& Hachisu (2004)]{kh04}           Kato, M., \& Hachisu, I. 2004, ApJ, 613, L129
\bibitem[Kato et al. (2008)]{kat08}              Kato, M., Hachisu I., Kiyota S., \& Saio, H. 2008, ApJ, 684, 1366
\bibitem[Li \& van den Heuvel (1997)]{lix97}     Li, X.-D., \& van den Heuvel, E.P.J. 1997, A\&A, 322, L9
\bibitem[Liu et al. (2010)]{liu10}               Liu, W.-M., Chen, W.-C., Wang, B., \& Han, Z. 2010, A\&A, 523, A3
\bibitem[Mannucci et al. (2006)]{man06}          Mannucci, F., Della Valle, M., \& Panagia, N. 2006, MNRAS, 370, 773
\bibitem[Meng \& Yang (2010)]{men10}             Meng, X., \& Yang, W. 2010, ApJ, 710, 1310
\bibitem[Mereghetti et al. (2009)]{mer09}        Mereghetti, S., Tiengo, A., Esposito, P., et al. 2009, Science, 325, 1222
\bibitem[Nomoto \& Iben (1985)]{nom85}           Nomoto, K., \& Iben, I. 1985, ApJ, 297, 531
\bibitem[Nomoto et al. (1997)]{nom97}            Nomoto, K., Iwamoto, K., \& Kishimoto, N. 1997, Science, 276, 1378
\bibitem[Wang \& Han (2009)]{wh09}               Wang, B., \& Han, Z. 2009, A\&A, 508, L27
\bibitem[Wang \& Han (2010a)]{wh10a}             Wang, B., \& Han, Z. 2010a, RAA (Res. Astron. Astrophys.), 10, 235
\bibitem[Wang \& Han (2010b)]{wh10b}             Wang, B., \& Han, Z. 2010b, MNRAS, 404, L84
\bibitem[Wang \& Han (2010c)]{wh10c}             Wang, B., \& Han, Z. 2010c, A\&A, 515, A88
\bibitem[Wang \& Han (2010d)]{wh10d}             Wang, B., \& Han, Z. 2010d, RAA (Res. Astron. Astrophys.), 10, 681
\bibitem[Wang, Li \& Han (2010)]{wlh10}          Wang, B., Li, X.-D., \& Han, Z. 2010, MNRAS, 401, 2729
\bibitem[Wang et al. (2010)]{wan10}              Wang, B., Liu, Z., Han, Y., et al. 2010, ScChG (Sci. China Ser. G),  53, 586
\bibitem[Wang et al. (2009a)]{wan09a}            Wang, B., Meng, X., Chen, X., \& Han, Z. 2009a, MNRAS, 395, 847
\bibitem[Wang et al. (2009b)]{wan09b}            Wang, B., Chen, X., Meng, X., \& Han, Z. 2009b, ApJ, 701, 1540
\bibitem[Webbink (1984)]{web84}                  Webbink, R.F. 1984, ApJ, 277, 355



\end{thebibliography}
\end{document}